\documentclass{article} 
\usepackage{nips12submit_e, times, amsthm, amsmath, bbm, amsfonts, amssymb, cite, graphicx, caption, enumerate, stmaryrd, ctable, subfigure}
\usepackage{hyperref,cancel}

\newcommand{\X}{{\mathcal{X}}}
\newcommand{\Y}{{\mathcal{Y}}}
\newcommand{\data}{{\mathcal D}}

\newcommand{\bE}{{\mathbb E}}

\newcommand{\bR}{{\mathbb R}}
\newcommand{\bI}{{\mathbb I}}
\newcommand{\cF}{{\mathcal F}}

\newcommand{\cL}{{\mathcal L}}
\newcommand{\fm}{{\mathfrak m}}

\newcommand{\wt}{{\mathbf w}}

\title{Regulating the information in spikes:\\a useful bias}

\author{David Balduzzi\\
Department of Computer Science\\
ETH Zurich, Switzerland\\
\texttt{david.balduzzi@inf.ethz.ch}}

\nipsfinalcopy 

\begin{document}
\maketitle

The bias/variance tradeoff is fundamental to learning: increasing a model's complexity can improve its fit on training data, but potentially worsens performance on future samples \cite{vapnik:82}. Remarkably, however, the human brain effortlessly handles a wide-range of complex pattern recognition tasks. On the basis of these conflicting observations, it has been argued that \emph{useful biases} in the form of ``generic mechanisms for representation'' must be hardwired into cortex \cite{geman:92}.

This note describes a useful bias that encourages cooperative learning which is both biologically plausible and rigorously justified  \cite{bt:08, balduzzi:11ffp, bb:12, bt:12, bob:12, Nere:2012fk, Tononi:06}. 

Let us outline the problem. Neurons learn inductively. They generalize from finite samples and encode estimates of future outcomes (for example, rewards) into their spiketrains \cite{Gottfried:2003fk}. Results from learning theory imply that generalizing successfully requires strong biases \cite{vapnik:82} or, in other words, specialization. Thus, at any given time some neurons' specialties are more relevant than others. Since most of the data neurons receive are other neurons' outputs, it is essential that neurons indicate which of their outputs encode high quality estimates. Downstream neurons should then be biased to specialize on these outputs.

The proposed biasing mechanism is based on a constraint on the effective information, $ei$, generated by spikes, see Eq.~\eqref{e:constraint} below. The motivation for using effective information comes from a connection to learning theory explained in \S\ref{s:learning}. There, we show the $ei$ generated by empirical risk minimization quantifies capacity: higher $ei$ yields tighter generalization bounds. 

Sections \S\ref{s:regularize} and \S\ref{s:cortex} consider implications of the constraint in two cases: abstractly and for a concrete model. In both cases we find that imposing constraint \eqref{e:constraint} implies: {\rm (i)} essentially all information is carried by spikes; {\rm (ii)} spikes encode reward estimates and {\rm (iii)} the higher the effective information, the better the guarantees on estimates. 

Although the proposal is inspired by cortical learning, the main ideas are  information-theoretic, suggesting they may also apply to other examples of interacting populations of adaptive agents.

\section{Information}
\label{s:information}

We model physical systems as input/output devices. For simplicity, we require that inputs $\X$ and outputs $\Y$ form finite sets. Systems are not necessarily deterministic. We encode the probability that system $\fm$ outputs $y\in \Y$ given input $x\in\X$ in Markov matrix $P_\fm(y|x)$. 

Consider two perspectives on a physical system. The first is \emph{computational}: the system receives an input, and we ask what output it will produce according to the probabilistic or deterministic rule encoded in $P_\fm$. The second is \emph{inferential}: the system produces an output, and we ask what information it generates about its input.

The inferential perspective is based on the notion of Bayesian information gain. Suppose we have model $P_{\mathcal M}(d|h)$ that specifies the probability of observing data given a hypothesis and also prior distribution $P(h)$ on hypotheses. If we  observe data $d$, how much have we learned about the hypotheses? The \textbf{Bayesian information gain} is 
\begin{equation}
	\label{e:big}
	D\big[P_{\mathcal M}(H|d)\,\big\|\,P(H)\big],
	\text{ where }D[P\|Q]:=\sum_i P_i\log_2\frac{P_i}{Q_i}	
\end{equation}
is the Kullback-Leibler divergence and $P_{\mathcal M}(h|d)$ is computed via Bayes' rule. 

Effective information is simply\footnote{In more general settings, effective information makes essential use of Pearl's interventional calculus \cite{pearl:00}.} the Bayesian information gain of physical system $\fm$, where inputs and outputs correspond to hypotheses and data respectively \cite{bt:08}. In the absence of additional constraints, we place the uniform (maximum entropy) prior on inputs. The \textbf{effective information} about $\X$ generated by $\fm$ when it outputs $y$ is 
\begin{equation}
	ei(\fm,y) := D\big[P_\fm(X|y)\,\big\|\,P_{unif}(X)\big].
	\label{e:ei}
\end{equation}
Effective information quantifies the distinctions in the input that are ``visible'' to $\fm$, insofar as they make a different to its output.

Finally, observe that effective information quantifies selectivity:
\begin{equation}
	\label{e:explanatory-power}
	ei(\fm, y) = \underbrace{H\big[P_{unif}(X)\big]}_{\text{total bits available}}\,\,\, - 
	\underbrace{H[P_\fm(X|y)]}_{\text{bits indistinguishable to }\fm}
	= \Big\{\text{selectivity of response $y$ by $\fm$}\Big\},
\end{equation}
where $H[\bullet]$ denotes entropy. Outputs generating higher effective information therefore trace back to more specific causes (i.e. more concentrated posteriors).

\section{Learning}
\label{s:learning}

The information one system generates about another should have implications for their future interactions. We show this holds for the well-studied special case of \textbf{empirical risk minimization} (ERM). Results are taken from \cite{balduzzi:11ffp}, which should be consulted for details.

Given function class $\cF\subset\Sigma_\X=\{\sigma:\X\rightarrow \pm1\}$ and unlabelled data $\data = (x_1,\ldots,x_\ell)\in\X^\ell$, ERM takes labelings of $\data$ to the empirical error of the best-performing classifier in $\cF$
\begin{equation}
	\label{e:erm}
	{\mathcal E}_{\cF,\data}:\Sigma_\X \longrightarrow \bR:\sigma\mapsto 
	\min_{f\in\cF}\frac{1}{\ell}\sum_{i=1}^\ell \bI_{f(x_i)\neq \sigma(x_i)}.
\end{equation}

It is easy to show that $ei({\mathcal E}_{\cF,\data},0) = \ell - VC_\cF(\data)$, where $VC_\cF(\data)$ is the \textbf{empirical VC-entropy} \cite{boucheron:00, balduzzi:11ffp}. It follows with high probability that
\begin{equation}
	\label{e:bound}
	\underbrace{\bE\left[\bI_{\widehat{f}(x)\neq \sigma(x)}\right]}_{\text{expected error}} 
	\leq \underbrace{\frac{1}{\ell}\sum_{i=1}^\ell\bI_{\widehat{f}(x_i)\neq \sigma(x_i)}}_{\text{training error}}
	+ c_1\sqrt{1-\underbrace{\frac{ei({\mathcal E}_{\cF,\data},0)}{\ell}}_{\text{effective information}}} 
	+ \Big\{\text{confidence term}\Big\}.
\end{equation}
Effective information answers the question: To what extent does the error take the form it does because of the supervisor? Note that the optimal classifier does not depend on the supervisor (``environment'') directly, but rather on the supervisor factored through ${\mathcal E}_{\cF,\data}$, which outputs the error:
\begin{equation}
	\cL_{\cF,\data}:\Sigma_\data\longrightarrow \cF:\sigma\mapsto
	\arg\underbrace{\min_{f\in\cF}\frac{1}{\ell}\sum_{i=1}^\ell \bI_{f(x_i)\neq \sigma(x_i)}}_{{\mathcal E}_{\cF,\data}(\sigma)}
	=: \widehat{f}
\end{equation}
A learning algorithm that shatters the data achieves zero empirical error for all supervisors, which implies the \emph{error is independent of the supervisor}. Equivalently, it implies the error achieved by ERM generates no information about the supervisor. Increasing the effective information generated by ${\mathcal E}_{\cF,\data}$ progressively concentrates the distribution of likely future errors. Guarantees tighten as $ei$ increases. 

An empirical risk minimizer's training error is a \emph{meaningful indicator of future performance} to the extent that it generates high effective information. This motivates investigating whether the effective information generated when performing more biologically plausible optimizations also has implications for future performance.

\section{Cooperative learning in abstract}
\label{s:regularize}

Organisms aim to choose beneficial actions in the situations they encounter. Responsibility for making choices falls largely on the cortex. However, cortical neurons interact extraordinarily indirectly with the external environment -- their inputs and outputs are mediated through millions of other neurons that constantly rewire themselves. It follows that, at best, neurons can encode provisional estimates of expected outcomes into their spiketrains. 

\emph{Guaranteeing and highlighting} high quality estimates of future outcomes is therefore essential. In particular, since interneuronal communication is dominated by spikes, it is necessary that spikes provide \emph{meaningful indicators} of future outcomes. 

We propose an information-theoretic constraint that simultaneously guarantees and highlights estimates in populations of interacting learners \cite{bb:12, bt:12, bob:12, Nere:2012fk}. The next section considers a particular model of cortical neurons in more detail. 

Let us model abstract learners as adaptable channels $X\xrightarrow{\fm}Y$ with two outputs, suggestively called spikes $y_1$ and silences $y_0$. Impose the following 
\begin{equation}
	\label{e:constraint}
	\tag{*}
	\text{\textbf{constraint: }}\,\,\,\,\,\, 
	ei(\fm,y_1) = \lambda\text{ bits},\,\,\,\,
	\text{ where }\lambda\gg ei(\fm,y_0).
\end{equation}
Parameter $\lambda$ controls the fraction of inputs causing the learner to produce $y_1$ (the higher $\lambda$, the smaller the fraction). It makes sense that each learner in a large population should specialize on a small fraction of possible inputs. Indeed, increasing $\lambda$ decreases the frequency of $y_1$, so $P_\fm(y_1)\ll P_\fm(y_0)$.

\paragraph{Consequence 1: Spikes carry (essentially all) information.}
Under constraint \eqref{e:constraint}, the information transferred by learners is carried by spikes alone
\begin{equation}
	\label{e:transfer}
	I(X;Y) = P_\fm(y_1)\cdot ei(\fm,y_1) + O\left(P_\fm(y_1)^2\right)
	\approx  P_\fm(y_1)\cdot ei(\fm,y_1)
\end{equation}
to a first-order approximation \cite{bob:12}.  It follows that silent learners can be ignored, despite the fact that they typically constitute the bulk of the population's responses (recall: $P_\fm(y_1)\ll P_\fm(y_0)$). 

This suggests a principled way to distribute credit \cite{bt:12}. When a positive/negative global signal is released, the few learners with informative responses -- that trace back to specific stimuli, recall \eqref{e:explanatory-power} -- should reinforce/weaken their behaviors. Conversely, the many learners producing uninformative responses that do not trace back to specific stimuli should not modify their synapses. It turns out this is exactly how neurons modify their synapses, see \S\ref{s:cortex}.

\paragraph{Consequence 2: Spikes encode high quality reward estimates.}
Under constraint \eqref{e:constraint}, the effective information generated by reward maximizing\footnote{formalized as a constrained optimization following the free-energy approaches developed in \cite{OrtegaBraun:11, tishby:11}.} learners essentially equals their empirical relative reward \cite{bob:12}. Thus, outputs with high $ei$ indicate high empirical reward relative to alternatives. 

Furthermore, constraint \eqref{e:constraint} ensures that the information encoded in spikes is reliable. Applying a PAC-Bayes inspired variant of Ockham's razor \cite{seldin:08}, it can be shown that the higher the effective information generated by spikes, the smaller the difference between the empirical reward estimate $\widehat{R}_\fm$ and expected reward $R_\fm$:
\begin{equation}
	\label{e:pac}
	\underbrace{\big|R_\fm - \widehat{R}_\fm\big|}_{\text{accuracy of reward estimate}}
	\leq \sqrt{\frac{c_1}{T_1}}
	\cdot\underbrace{\sqrt{\frac{ei(\fm,y_1)+1}{\exp[ei(\fm,y_1)]}
	+ \Big\{\text{confidence term}\Big\}}}_{\text{term that decreases as }ei\text{ increases}}.
\end{equation}

\paragraph{Summary.}
The information-theoretic constraint \eqref{e:constraint} introduces an asymmetry into outputs, ensures that spikes carry reliable information about future rewards.

\section{Cooperative learning in cortex}
\label{s:cortex}

Spike-timing dependent plasticity (STDP) is a standard and frequently extended model of plasticity \cite{song:00}. Unfortunately, it operates in continuous time and is difficult to analyze using standard learning-theoretic techniques. Recently \cite{bb:12}, we investigated the fast-time constant limit of STDP. Taking the limit strips out the exponential discount factors, essentially reducing STDP to a simpler discrete time algorithm whilst preserving its essential structure.

We recall STDP. Given a presynaptic spike at time $t_j$ and postsynaptic at time $t_k$, the strength of synapse $j\rightarrow k$ is updated according to
\begin{equation}	
	\label{e:stdp}
	\Delta \wt_{jk} =\begin{cases}
 	\alpha_+\cdot e^{\left(\frac{t_j-t_k}{\tau_+}\right)} & \mbox{ if }t_j\leq t_k
	\\  
	-\alpha_-\cdot e^{\left(\frac{t_k-t_j}{\tau_-}\right)} & \mbox{ else}.
	\end{cases}
\end{equation}
STDP modifies synapses when input \emph{and} output spikes co-occur in a short time window, so that spikes gate learning. This makes sense if spikes are selective, and so by \eqref{e:transfer} carry most of the information in cortex.

\paragraph{Synaptic weights control expected error.}
If STDP incorporates neuromodulatory signals then it can be shown to encode reward estimates into spikes in the fast-time constant limit $\tau_\bullet\rightarrow0$, see \cite{bb:12}.  The accuracy of these estimates is controlled by synaptic weights.

More precisely, let $\nu:X\rightarrow \bR$ be a random variable representing neuromodulatory signals; $l$ denote the error, i.e. whether a spike is followed by a negative neuromodulatory signal; and $R$ the empirical reward after spiking \cite{bb:12}. The expected error is controlled by the sum $\omega$ of neuron $\fm$'s synaptic weights and $\fm$'s empirical reward: 
\begin{equation}
	\label{e:rew-error}
	\underbrace{\bE\big[l(x,\fm,\nu)\big]}_{\text{expected error}}
	\leq c_1\cdot\omega^2
	- \underbrace{\sum_i R\big(x^{(i)},\fm,\nu\big)}_{\text{empirical reward}}
	+\,\, \omega\cdot\Big\{\text{capacity term}\Big\}
	+ \Big\{\text{confidence term}\Big\}.
\end{equation}
Thus, \emph{the lower synaptic weights $\omega$, the better the quality of spikes as indicators of future outcomes.}

\paragraph{Synaptic weights are homeostatically regulated.}
Spikes are both metabolically expensive \cite{Hasenstaub:2010fk} and selective: they typically occur in response to specific stimuli -- e.g. an edge or a familiar face \cite{quiroga:05}. Using spikes selectively reduces metabolic expenditures, which is important since the brain accounts for a disproportionate fraction of the body's total energy budget \cite{attwell:01}.

There is evidence that  synaptic strengths increase on average during wakefulness and are downscaled during sleep \cite{Tononi:06, vyazovskiy:08, Maret:2011rt}. This suggests that homeostatic regulation during sleep may both reduce metabolic costs and simultaneously improve guarantees on reward estimates. Finally, it is easy to show that \emph{decreasing synaptic weights increases the effective information generated by spikes}.

\section{Discussion}

There is an interesting analogy between spikes and paper currency. Money plays many overlapping roles in an economy, including: {\rm (i)} focusing attention; {\rm (ii)} stimulating activity; and {\rm (iii)} providing a quantitative lingua franca for tracking revenues and expenditures. 

Spikes may play similar roles in cortex. Spikes focus attention: STDP and other proposed learning rules are particularly sensitive to spikes and spike timing. Spikes stimulate activity: input spikes cause output spikes. Finally, spikes leave trails of (Calcium) traces that are used to reinforce and discourage neuronal behaviors in response to neuromodulatory signals. 

Neither money nor spikes are intrinsically valuable. Currency can be devalued by inflation. Similarly, the information content and guarantees associated with spikes are eroded by overpotentiating synapses which reduces their selectivity (potentially leading to epileptic seizures in extreme cases). Regulating the information content of spikes is therefore essential. Eq.~\eqref{e:constraint} provides a simple constraint that can be approximately imposed by regulating synaptic weights. Indeed, there is evidence that one of the functions of sleep is precisely this. 

Spikes with high information content are valuable because they come with strong guarantees on their estimates. They are therefore worth paying attention to, worth responding to, worth keeping track of, and worth learning from.

\paragraph{Acknowledgements.} I thank Michel Besserve, Samory Kpotufe, Pedro Ortega for useful discussions. 

{\small

\newcommand{\BMCxmlcomment}[1]{}

\BMCxmlcomment{

<refgrp>

<bibl id="B1">
  <title><p>Estimation of {D}ependencies {B}ased on {E}mpirical
  {D}ata</p></title>
  <aug>
    <au><snm>Vapnik</snm><fnm>V</fnm></au>
  </aug>
  <publisher>Springer</publisher>
  <pubdate>1982</pubdate>
</bibl>

<bibl id="B2">
  <title><p>Neural {N}etworks and the {B}ias/{V}ariance {D}ilemma</p></title>
  <aug>
    <au><snm>Geman</snm><fnm>S</fnm></au>
    <au><snm>Bienenstock</snm><fnm>E</fnm></au>
    <au><snm>Doursat</snm><fnm>R</fnm></au>
  </aug>
  <source>Neural Comp</source>
  <pubdate>1992</pubdate>
  <volume>4</volume>
  <fpage>1</fpage>
  <lpage>-58</lpage>
</bibl>

<bibl id="B3">
  <title><p>Integrated {I}nformation in {D}iscrete {D}ynamical {S}ystems:
  {M}otivation and {T}heoretical {F}ramework.</p></title>
  <aug>
    <au><snm>Balduzzi</snm><fnm>D</fnm></au>
    <au><snm>Tononi</snm><fnm>G</fnm></au>
  </aug>
  <source>PLoS Comput Biol</source>
  <publisher>Department of Psychiatry, University of Wisconsin, Madison,
  Wisconsin, USA.</publisher>
  <pubdate>2008</pubdate>
  <volume>4</volume>
  <issue>6</issue>
  <fpage>e1000091</fpage>
</bibl>

<bibl id="B4">
  <title><p>Falsification and {F}uture {P}erformance</p></title>
  <aug>
    <au><snm>Balduzzi</snm><fnm>D</fnm></au>
  </aug>
  <source>Proceedings of Solomonoff 85th Memorial Conference</source>
  <publisher>Springer</publisher>
  <series><title><p>LNAI</p></title></series>
  <pubdate>in press</pubdate>
</bibl>

<bibl id="B5">
  <title><p>Towards a learning-theoretic analysis of spike-timing dependent
  plasticity</p></title>
  <aug>
    <au><snm>Balduzzi</snm><fnm>D</fnm></au>
    <au><snm>Besserve</snm><fnm>M</fnm></au>
  </aug>
  <source>To appear in Adv Neural Information Processing Systems
  (NIPS)</source>
  <pubdate>2012</pubdate>
</bibl>

<bibl id="B6">
  <title><p>What can neurons do for their brain? {C}ommunicate selectivity with
  spikes</p></title>
  <aug>
    <au><snm>Balduzzi</snm><fnm>D</fnm></au>
    <au><snm>Tononi</snm><fnm>G</fnm></au>
  </aug>
  <source>To appear in Theory in Biosciences</source>
  <pubdate>2012</pubdate>
</bibl>

<bibl id="B7">
  <title><p>Metabolic cost as an organizing principle for cooperative
  learning</p></title>
  <aug>
    <au><snm>Balduzzi</snm><fnm>D</fnm></au>
    <au><snm>Ortega</snm><fnm>PA</fnm></au>
    <au><snm>Besserve</snm><fnm>M</fnm></au>
  </aug>
  <source>Under review, \href{http://xxx.lanl.gov/abs/1202.4482}</source>
  <pubdate>2012</pubdate>
</bibl>

<bibl id="B8">
  <title><p>A neuromorphic architecture for object recognition and motion
  anticipation using burst-{S}{T}{D}{P}</p></title>
  <aug>
    <au><snm>Nere</snm><fnm>A</fnm></au>
    <au><snm>Olcese</snm><fnm>U</fnm></au>
    <au><snm>Balduzzi</snm><fnm>D</fnm></au>
    <au><snm>Tononi</snm><fnm>G</fnm></au>
  </aug>
  <source>PLoS One</source>
  <pubdate>2012</pubdate>
  <volume>7</volume>
  <issue>5</issue>
  <fpage>e36958</fpage>
</bibl>

<bibl id="B9">
  <title><p>Sleep function and synaptic homeostasis.</p></title>
  <aug>
    <au><snm>Tononi</snm><fnm>G</fnm></au>
    <au><snm>Cirelli</snm><fnm>C</fnm></au>
  </aug>
  <source>Sleep Med Rev</source>
  <publisher>Department of Psychiatry, University of Wisconsin, 6001 Research
  Park Blvd., Madison, WI 53719, USA. gtononi@wisc.edu</publisher>
  <pubdate>2006</pubdate>
  <volume>10</volume>
  <issue>1</issue>
  <fpage>49</fpage>
  <lpage>-62</lpage>
</bibl>

<bibl id="B10">
  <title><p>Encoding predictive reward value in human amygdala and
  orbitofrontal cortex</p></title>
  <aug>
    <au><snm>Gottfried</snm><fnm>JA</fnm></au>
    <au><snm>O'Doherty</snm><fnm>J</fnm></au>
    <au><snm>Dolan</snm><fnm>RJ</fnm></au>
  </aug>
  <source>Science</source>
  <pubdate>2003</pubdate>
  <volume>301</volume>
  <issue>5636</issue>
  <fpage>1104</fpage>
  <lpage>7</lpage>
</bibl>

<bibl id="B11">
  <title><p>Causality: models, reasoning and inference</p></title>
  <aug>
    <au><snm>Pearl</snm><fnm>J</fnm></au>
  </aug>
  <publisher>Cambridge University Press</publisher>
  <pubdate>2000</pubdate>
</bibl>

<bibl id="B12">
  <title><p>A {S}harp {C}oncentration {I}nequality with
  {A}pplications</p></title>
  <aug>
    <au><snm>Boucheron</snm><fnm>S</fnm></au>
    <au><snm>Lugosi</snm><fnm>G</fnm></au>
    <au><snm>Massart</snm><fnm>P</fnm></au>
  </aug>
  <source>Random Structures and Algorithms</source>
  <pubdate>2000</pubdate>
  <volume>16</volume>
  <issue>3</issue>
  <fpage>277</fpage>
  <lpage>292</lpage>
</bibl>

<bibl id="B13">
  <title><p>Information, utility and bounded rationality</p></title>
  <aug>
    <au><snm>Ortega</snm><fnm>PA</fnm></au>
    <au><snm>Braun</snm><fnm>DA</fnm></au>
  </aug>
  <source>The fourth conference on artificial general intelligence</source>
  <pubdate>2011</pubdate>
  <fpage>269</fpage>
  <lpage>274</lpage>
</bibl>

<bibl id="B14">
  <title><p>Information theory of decisions and actions</p></title>
  <aug>
    <au><snm>Tishby</snm><fnm>N</fnm></au>
    <au><snm>Polani</snm><fnm>D</fnm></au>
  </aug>
  <source>Perception-{R}eason-{A}ction</source>
  <publisher>Springer</publisher>
  <editor>Vassilis and Hussain and Taylor</editor>
  <pubdate>2011</pubdate>
</bibl>

<bibl id="B15">
  <title><p>Multi-{C}lassification by {C}ategorical {F}eatures via
  {C}lustering</p></title>
  <aug>
    <au><snm>Seldin</snm><fnm>Y</fnm></au>
    <au><snm>Tishby</snm><fnm>N</fnm></au>
  </aug>
  <source>Proceedings of the 25th {I}nternational {C}onference on {M}achine
  {L}earning</source>
  <pubdate>2008</pubdate>
</bibl>

<bibl id="B16">
  <title><p>Competitive {H}ebbian learning through spike-timing-dependent
  synaptic plasticity</p></title>
  <aug>
    <au><snm>Song</snm><fnm>S</fnm></au>
    <au><snm>Miller</snm><fnm>K D</fnm></au>
    <au><snm>Abbott</snm><fnm>L F</fnm></au>
  </aug>
  <source>Nature Neuroscience</source>
  <pubdate>2000</pubdate>
  <volume>3</volume>
  <issue>9</issue>
</bibl>

<bibl id="B17">
  <title><p>Metabolic cost as a unifying principle governing neuronal
  biophysics</p></title>
  <aug>
    <au><snm>Hasenstaub</snm><fnm>A</fnm></au>
    <au><snm>Otte</snm><fnm>S</fnm></au>
    <au><snm>Callaway</snm><fnm>E</fnm></au>
    <au><snm>Sejnowski</snm><fnm>TJ</fnm></au>
  </aug>
  <source>Proc Natl Acad Sci U S A</source>
  <pubdate>2010</pubdate>
  <volume>107</volume>
  <issue>27</issue>
  <fpage>12329</fpage>
  <lpage>34</lpage>
</bibl>

<bibl id="B18">
  <title><p>Invariant visual representation by single neurons in the human
  brain</p></title>
  <aug>
    <au><snm>Quiroga</snm><fnm>R Q</fnm></au>
    <au><snm>Reddy</snm><fnm>L</fnm></au>
    <au><snm>Koch</snm><fnm>C</fnm></au>
    <au><snm>Fried</snm><fnm>I</fnm></au>
  </aug>
  <source>Nature</source>
  <pubdate>2005</pubdate>
  <volume>435</volume>
  <issue>1102-1107</issue>
</bibl>

<bibl id="B19">
  <title><p>An energy budget for signaling in the grey matter of the
  brain</p></title>
  <aug>
    <au><snm>Attwell</snm><fnm>D</fnm></au>
    <au><snm>Laughlin</snm><fnm>S B</fnm></au>
  </aug>
  <source>J Cereb Blood Flow Metab</source>
  <pubdate>2001</pubdate>
  <volume>21</volume>
  <issue>10</issue>
  <fpage>1133</fpage>
  <lpage>45</lpage>
</bibl>

<bibl id="B20">
  <title><p>Molecular and electrophysiological evidence for net synaptic
  potentiation in wake and depression in sleep</p></title>
  <aug>
    <au><snm>Vyazovskiy</snm><fnm>V V</fnm></au>
    <au><snm>Cirelli</snm><fnm>C</fnm></au>
    <au><snm>Pfister Genskow</snm><fnm>M</fnm></au>
    <au><snm>Faraguna</snm><fnm>U</fnm></au>
    <au><snm>Tononi</snm><fnm>G</fnm></au>
  </aug>
  <source>Nat Neurosci</source>
  <pubdate>2008</pubdate>
  <volume>11</volume>
  <issue>2</issue>
  <fpage>200</fpage>
  <lpage>8</lpage>
</bibl>

<bibl id="B21">
  <title><p>Sleep and waking modulate spine turnover in the adolescent mouse
  cortex.</p></title>
  <aug>
    <au><snm>Maret</snm><fnm>S</fnm></au>
    <au><snm>Faraguna</snm><fnm>U</fnm></au>
    <au><snm>Nelson</snm><fnm>AB</fnm></au>
    <au><snm>Cirelli</snm><fnm>C</fnm></au>
    <au><snm>Tononi</snm><fnm>G</fnm></au>
  </aug>
  <source>Nat Neurosci</source>
  <publisher>Department of Psychiatry, University of Wisconsin, Madison,
  Wisconsin, USA.</publisher>
  <pubdate>2011</pubdate>
  <volume>14</volume>
  <issue>11</issue>
  <fpage>1418</fpage>
  <lpage>-1420</lpage>
</bibl>

</refgrp>
} 
}

\end{document}